\documentstyle[emulateapj,psfig]{article}

\newcommand{\be}{\begin{equation}}
\newcommand{\ee}{\end{equation}}
\newcommand{\ba}{\begin{eqnarray}}
\newcommand{\ea}{\end{eqnarray}}
\newcommand{\siml}{\lower4pt \hbox{$\buildrel < \over \sim$}}
\newcommand{\simg}{\lower4pt \hbox{$\buildrel > \over \sim$}}
\def\Mesz{M\'esz\'aros}

\slugcomment{ApJ in press}

\begin{document}

\title{Gamma-ray burst early optical afterglows: implications for
the initial Lorentz factor and the central engine}

\author{Bing Zhang$^1$, Shiho Kobayashi$^{1,2}$, \& Peter
M\'esz\'aros$^{1,2}$}
\affil{
$^{1}$Department of Astronomy \& Astrophysics, Pennsylvania State
University, University Park, PA 16802 \\
$^{2}$Deptartment of Physics, Pennsylvania State University,
University Park, PA 16802}

\begin{abstract}
Early optical afterglows have been observed from GRB 990123, GRB
021004, and GRB 021211, which reveal rich emission features attributed
to reverse shocks. It is expected that {\em Swift} will discover
many more early afterglows. Here we investigate in a unified manner
both the forward and the reverse external shock emission components,
and introduce a straightforward recipe for directly constraining the
initial Lorentz factor of the fireball using early optical afterglow
data. The scheme is largely independent of the shock microphysics. We
identify two types of combinations of the reverse and forward shock
emission, and explore their parameter regimes. We also discuss a
possible diagnostic for magnetized ejecta. There is evidence that the
central engine of GRB 990123 is strongly magnetized.

\end{abstract}


\section{Introduction}
The standard Gamma-ray burst (GRB) afterglow model (\Mesz~\& Rees
1997a; Sari, Piran \& Narayan 1998) invokes synchrotron emission of
electrons from the forward external shock, and has been proven
successful in interpreting the late time broadband afterglows. At these
late times the fireball is already decelerated and has entered a
self-similar regime, in which precious information about the early
ultra-relativistic phase is lost. In the very early afterglow epoch,
the emission from the reverse shock propagating into the fireball
itself also plays a noticeable role, especially in the low frequency
bands, e.g. optical or radio (\Mesz~\& Rees 1997a; Sari \& Piran 1999b),
and information about the fireball initial Lorentz factor could be in
principle retrieved from the reverse shock data. For a long time,
evidence for reverse shock emission was available only from GRB 990123
(Akerlof et al. 1999; Sari \& Piran 1999a; \Mesz~\& Rees 1999; Kobayashi
\& Sari 2000). Recently, thanks to prompt localizations of GRBs by the
{\em High Energy Transient Explorer 2 (HETE-2)} (e.g. Shirasaki et al.
2002; Crew et al. 2002) and rapid follow-ups by robotic optical
telescopes (e.g. Fox 2002; Li et al. 2002; Fox \& Price 2002; Park et
al. 2002; Wolzniak et al. 2002), reverse shock emission has also been
identified from GRB 021211 (Fox et al. 2003; Li et al. 2003; Wei
2003) and possibly also from GRB 021004 (Kobayashi \& Zhang 2003a). It
is expected that the {\em Swift} mission will
record many GRB early optical afterglows after its launch scheduled in
December 2003, which will unveil a rich phenomenology of early
afterglows.

Here we propose a paradigm to analyze the early optical afterglow data,
starting from tens of seconds after the gamma-ray trigger. By combining
the emission information from both the forward and the reverse shocks,
we discuss a way to derive or constrain the initial Lorentz factor of
the fireball directly from the observables. The method does not depend
on the absolute values of the poorly-known shock microphysics
(e.g. the electron and magnetic equipartition parameters $\epsilon_e$
and $\epsilon_B$, and the electron power-law index $p$). We
also categorize the early optical afterglows into two types and
discuss the parameter regimes for both cases.

\section{Forward and reverse shock comparison}

We consider a relativistic shell (fireball ejecta) with an
isotropic equivalent energy $E$ and an initial Lorentz factor
$\gamma_0$ expanding into a homogeneous interstellar medium (ISM)
with particle number density $n$ at a redshift $z$. Some GRB
afterglows may occur in the progenitors' stellar winds (e.g.
Chevalier \& Li 1999). The combined reverse vs. forward shock
emission for the wind environment is easy to be distinguished from
what is discussed here, and has been discussed separately
(Kobayashi \& Zhang 2003b).

For the homogeneous ISM case, in the observer's frame, one can define
a timescale when the accumulated ISM mass is $1/\gamma_0$ of the
ejecta mass, i.e., $t_\gamma=[(3E/4\pi \gamma_0^2 n m_p c^2)^{1/3} /2
\gamma_0^2 c] (1+z)$. This is the fireball deceleration time if the
burst duration $T < t_\gamma$ (the so-called thin-shell regime,
Sari \& Piran 1995). For $T > t_\gamma$, the deceleration
time is delayed to $T$ (thick-shell regime). The critical condition
$T=t_\gamma$ defines a critical initial Lorentz factor
\be
\gamma_c \simeq  125 E_{52}^{1/8} n^{-1/8} T_2^{-3/8}
\left(\frac{1+z}{2}\right)^{3/8}~,
\label{etac}
\ee
where the convention $Q=10^x Q_x$ is used. So one has $\gamma_0 <
\gamma_c$ for the thin-shell case, and $\gamma_0 > \gamma_c$ for the
thick-shell case. When the reverse shock
crosses the shell, the observer time and the ejecta Lorentz factor are
(Sari \& Piran 1995; Kobayashi, Piran \& Sari 1999)
\ba
t_\times = {\rm max} (t_\gamma, T), &
\gamma_\times = {\rm min} (\gamma_0, \gamma_c),
\label{tcross}
\ea
where the first and second values in the above expressions correspond
to the thin and thick shell cases, respectively.
Since $T$ and $z$ can be directly measured, e.g.
by {\em Swift}, and $E$ and $n$ could be inferred from broadband afterglow
fitting (e.g. Panaitescu \& Kumar 2001), $\gamma_c$ is essentially an
observable in an idealized observational campaign. Some
estimates (or fits) about the microphysics parameters ($\epsilon_e$,
$\epsilon_B$, $p$, etc.) have to be involved when inferring $E$ and
$n$, but $\gamma_c$ is insensitive to the inaccuracy of measuring $n$
and $E$. The correction to $\gamma_c$ is less than a factor of 2 even
if the uncertainty of either $E$ or $n$ is by a factor of 100.

The forward shock synchrotron spectrum can be approximated as a
four-segment power-law with breaks at the cooling frequency
$\nu_c$, typical frequency $\nu_m$ and the self-absorption
frequency $\nu_a$ (Sari et al. 1998). So is the reverse shock
emission spectrum at $t<t_\times$. For $t>t_\times$, there is
essentially no emission above $\nu_c$ for the reverse shock
emission. For both shocks, the typical frequency, cooling
frequency and the peak flux scale as $\nu_m \propto \gamma B
\gamma_e^2$, $\nu_c \propto \gamma^{-1} B^{-3} t^{-2}$, and
$F_{\nu,m} \propto \gamma B N_e$, where $\gamma$ is the bulk
Lorentz boost, $B$ is the comoving magnetic field, $\gamma_e$ is
the typical electron Lorentz factor in the shock-heated region,
and $N_e$ is the total number of emitting electrons. Using similar
analyses in Kobayashi \& Zhang 2003a, we can derive
\ba
\frac{\nu_{m,r}(t_\times)}{\nu_{m,f}(t_\times)} &\sim& \hat\gamma^{-2}
{\cal R}_B~,
\label{fnum}\\
\frac{\nu_{c,r}(t_\times)}{\nu_{c,f}(t_\times)} & \sim &
{{\cal R}_B}^{-3}~,
\label{fnuc}\\
\frac{F_{\nu,m,r}(t_\times)}{F_{\nu,m,f}(t_\times)}
&\sim& \hat\gamma {\cal R}_B ~, \label{fFnu}
\ea
where
\ba
\hat\gamma\equiv \frac{\gamma_\times^2}{\gamma_0} = {\rm min} (\gamma_0,
\frac{\gamma_c^2}{\gamma_0}) \leq \gamma_c~, \\
{\cal R}_B\equiv
\frac{B_r}{B_f}=\left(\frac{\epsilon_{B,r}}{\epsilon_{B,f}}\right)^{1/2}
~, \label{b}
\ea
and the subcripts `f' and `r' indicate forward and reverse shock,
respectively. By deriving the second equation in eq.(\ref{b}), we have 
taken into account the fact that the internal energy densities are the 
same in both the forward- and reverse-shocked regions, as evident in
the standard hydrodynamical shock jump conditions. We have also
assumed that the $\epsilon_e$ and 
$p$ parameters are the same for both the forward and reverse-shocked
regions\footnote{If these parameters are different, there will be an
additional factor $({\cal R}_e {\cal R}_g)^{2}$ on the right hand
side of eq.[\ref{fnum}], where ${\cal
R}_e=\epsilon_{e,r}/\epsilon_{e,f}$, ${\cal R}_g=g_r/g_f$,
and $g=(p-2)/(p-1)$. If indeed ${\cal R}_e$ and ${\cal R}_g$ do not
equal to unity, our whole discussion in the text can be modified
straightforwardly.},
but with different $\epsilon_B$'s (as parameterized by ${\cal R}_B$). The
reason we introduce the ${\cal R}_B$ parameter is that in
some central engine models (e.g. Usov 1992; \Mesz~\& Rees 1997b;
Wheeler et al. 2000),
the fireball wind may be endowed with ``primordial'' magnetic fields,
so that in principle $B_r$ could be higher than $B_f$. We note that
all the previous investigations have assumed the same microphysics
parameters in both shocks.

The forward shock emission is likely to be in the
``fast cooling'' regime ($\nu_{c,f}<\nu_{m,f}$) initially and
in the ``slow cooling'' regime ($\nu_{c,f}>\nu_{m,f}$) at later times
(Sari et al. 1998). For the reverse shock emission, it can be deduced
from eqs. (\ref{fnum}, \ref{fnuc}) and eq.(11) of Sari et al. (1998)
that slow cooling ($\nu_{c,r} > \nu_{m,r}$) is generally valid as long
as ${\cal R}_B$ is not much greater than unity.

The lightcurves can be derived by specifying the temporal evolution
of $\nu_m$, $\nu_c$, and $F_{\nu,m}$. To first order, in the forward
shock (\Mesz ~\& Rees 1997a) we have
\ba
\nu_{m,f} \propto t^{-3/2}, & F_{\nu,m,f} \propto t^0
\label{f}
\ea
while in the reverse shock for $t>t_\times$ (Kobayashi 2000) we have
approximately
\ba
\nu_{m,r} \propto t^{-3/2}, & F_{\nu,m,r} \propto t^{-1}
\label{r}
\ea
The optical-band lightcurve for the forward shock is
well-described by the ``low-frequency'' case of Sari et al. (1998,
their Fig.2b), characterized by a turnover of the temporal indices
from 1/2 to $\sim -1$
at the peak time $t_{p,f}$ (at which $\nu_{m,f}$ crosses the observational
band, e.g. the typical R-band frequency $\nu_R$).
The optical lightcurve for the reverse shock emission is more
complicated, depending on whether the shell is thin or thick, whether
it is in the slow or fast cooling regime, and how $\nu_R$ compares with
$\nu_{m,r}(t_\times)$ and $\nu_{c,r}(t_\times)$ (see Kobayashi 2000
for a complete discussion). For reasonable parameters, however, the
cases of $\nu_R > \nu_{c,r}$ and $\nu_R < \nu_{a,r}$ are unlikely,
and even if these conditions are satisfied, they do not produce
interesting reverse shock signatures. Defining
\be
{\cal R}_\nu \equiv \frac{\nu_R}{\nu_{m,r}(t_\times)},
\label{zeta}
\ee
and specifying the slow-cooling case (which is reasonable for a large
sector of parameter regimes, see Kobayashi 2000), the reverse shock
lightcurves are then categorized into four cases with various temporal
indices separated by various break times.
The four cases with their typical temporal power-law indices
(in parentheses), which correspond to the thick and thin solid lines
in Fig.2a and Fig.3a of Kobayashi (2000), are:
(i) thin shell ${\cal R}_\nu > 1$: $\sim (5, -2)$;
(ii) thin shell ${\cal R}_\nu < 1$: $\sim (5, -1/2, -2)$:
(iii) thick shell ${\cal R}_\nu > 1$: $\sim(1/2, -2)$;
(iv) thick shell ${\cal R}_\nu < 1$: $\sim(1/2, -1/2, -2)$.
We note that in all four cases, the final temporal power-law index is
$\sim -2$, corresponding to adiabatic cooling of the already
accelerated electrons at $t_\times$, with $\nu_R >
\nu_{m,r}$. Strictly speaking, by writing the final lightcurve as
$F_{\nu,r}=F_{\nu,m,r}(\nu/\nu_{m,r})^{-(p-1)/2} \propto
t^{-\alpha}$, and making use of eq.(\ref{r}), we have
\be
\alpha = \frac{3p+1}{4},
\label{alpha}
\ee
where $p$ is the power law index of the electrons in the reverse shock
region. We can see that $\alpha \sim 2$ for typical $p$ values. Going
backwards in time, this well-known $F_{\nu,r} \propto t^{-2}$ reverse
shock lightcurve ends at $t_\times$ for ${\cal R}_\nu > 1$ (with a rising
lightcurve before that), but at
the time when $\nu_{m,r}$ crosses $\nu_R$ for ${\cal R}_\nu < 1$ (with a
flatter declining lightcurve, $F_{\nu,r} \propto t^{-1/2}$, before)
(Fig.1).

We define as $t_{p,r}$ the time when the $F_{\nu,r} \propto
t^{-2}$ lightcurve starts and define the corresponding reverse
shock spectral flux as $F_{\nu,p,r}$ (Fig.1). The lightcurve point
$(t_{p,r}, F_{\nu,p,r})$ is generally called the reverse shock
peak, but strictly this is valid only when ${\cal R}_\nu > 1$ (in
which case $t_{p,r}=t_\times$). On the other hand, it is natural
to define the forward shock peak at $(t_{p,f}, F_{\nu,p,f})$
(Fig.1), where $F_{\nu,p,f}=F_{\nu,m,f}=$constant (eq.[\ref{f}]).
For practical purposes, it is illustrative to derive the ratio of
the two peak times and that of the two peak fluxes by making use
of eqs. (\ref{fnum}), (\ref{fFnu}), (\ref{f}), (\ref{r}) and
(\ref{alpha}), i.e.
\begin{equation}
{{\cal R}_t} \equiv \frac{t_{p,f}}{t_{p,r}}=
\left\{
\begin{array}{@{\,}ll}
\hat\gamma^{4/3} {{\cal R}_B}^{-2/3} {\cal R}_\nu^{-2/3}, & {\cal
R}_\nu > 1, \\
\hat\gamma^{4/3} {{\cal R}_B}^{-2/3}, & {\cal R}_\nu < 1,
\end{array}
\right.
\label{ft}
\end{equation}
\begin{equation}
{{\cal R}_F}\equiv \frac{F_{\nu,p,r}}{F_{\nu,p,f}}=
\left\{
\begin{array}{@{\,}ll}
\hat\gamma {\cal R}_B {\cal R}_\nu^{-2(\alpha-1)/3},      & {\cal
R}_\nu > 1, \\
\hat\gamma {\cal R}_B {\cal R}_\nu^{2/3},      & {\cal R}_\nu < 1.
\end{array}
\right.
\label{fF}
\end{equation}
Notice that we have intentionally defined both ${{\cal R}_t}$ and
${{\cal R}_F}$ to be (usually) larger than unity, and that these
expressions are valid regardless of whether the shell is thin or
thick. Since we are working on the ``ratios'' of the quantities for
both shocks, the absolute values of the microphysics parameters do not
enter the problem. Instead, our expressions involve relative ratios of
these parameters, so that they cancel out in eqs.(\ref{ft})
and (\ref{fF}) if they are the same in both shocks, as we have
assumed in this paper (except for ${\cal R}_B$).

\section{A recipe to constrain the initial Lorentz factor}

Reverse shock emission data have been used to estimate the initial
Lorentz factor of GRB 990123 (Sari \& Piran 1999a; Kobayashi \&
Sari 2000; Wang, Dai \& Lu 2000; Soderberg \& Ramirez-Ruiz 2002;
Fan et al. 2002), but these studies depend on poorly-known 
shock microphysics parameters such as, e.g., $\epsilon_e$,
$\epsilon_B$, $p$, etc. Here we introduce a simple method which avoids
or at least reduces the dependence on such parameters.

We first consider an ``idealized'' observational campaign, which
may be realized with {\em Swift} or a similar facility. We assume
that the optical lightcurve is monitored starting early enough
that both the reverse and forward shock ``peaks'', i.e.,
$(t_{p,r}, F_{\nu,p,r})$ and $(t_{p,f}, F_{\nu,p,f})$ are
identified, and that the fast-decaying reverse shock lightcurve
index $\alpha$ is measured.  As a result, ${{\cal R}_t}$, ${{\cal
R}_F}$ and $\alpha$ are all known parameters. 
Since our method is most useful when both peaks are measured,
{\em we strongly recommend that the Swift UVOT instrument closely 
follow GRB early lightcurves until the forward shock peak is identified.} 
With eqs.(\ref{ft}) and
(\ref{fF}), we can directly solve for $\hat\gamma$ and ${\cal
R}_\nu$ for two ${\cal R}_\nu$ regimes, in terms of ${{\cal
R}_t}$, ${{\cal R}_F}$ and $\alpha$, as well as a free parameter
${\cal R}_B$ (which is conventionally taken as unity). For ${\cal
R}_\nu>1$ (in which case we see a rising lightcurve before
$t_{p,r}$), we have
\ba
\hat\gamma & = & \left (\frac{ {{\cal
R}_t}^{(\alpha-1)}{{\cal R}_B}^{(2\alpha+1)/3}}{{{\cal
R}_F}}\right )^{\frac{3}{4\alpha-7}},  \\
{\cal R}_\nu & = & \left ( \frac{{{\cal R}_t}^{3/2}
{{\cal R}_B}^3}{{{\cal R}_F}^2} \right )
^{\frac{3}{4\alpha-7}},
\label{zeta1}
\ea
and for ${\cal R}_\nu < 1$ (in which case we see a $\propto t^{-1/2}$
decaying lightcurve before $t_{p,r}$), we have
\ba
\hat\gamma & = & {{\cal R}_t}^{3/4} {{\cal R}_B}^{1/2}, \\
{\cal R}_\nu & = & \left ( \frac{ {{\cal R}_t}^{3/2} {{\cal R}_B}^3}{
{{\cal R}_F}^2} \right )^{-3/4}.
\label{zeta2}
\ea
Finally, the initial Lorentz factor $\gamma_0$ can be determined from
$\hat\gamma$ through
\begin{equation}
\gamma_0=
\left\{
\begin{array}{@{\,}ll}
\hat\gamma, & {\rm thin~shell} \\
\gamma_c^2/\hat\gamma, & {\rm thick~shell}
\end{array}
\right.
\label{eta}
\end{equation}
Since $\gamma_c$ is essentially a known parameter (and in the
thin-shell case $\gamma_0$ is independent of $\gamma_c$),
we can derive two $\gamma_0$ values (except for an unknown
parameter ${\cal R}_B$) directly from the data, which correspond
to the thin and thick shell cases, respectively. When $t_\times$
(which is usually the earliest transition point of the rising
lightcurve and the falling lightcurve) is measured, we can 
disentangle whether the shell is thin or thick by comparing
$t_\times$ with $T$, i.e., $T \sim t_\times$ corresponds to the
thick shell case, while $T < t_\times$ corresponds to the thin
shell case.

In reality, due to delay of telescope response, the reverse shock
peak time might not be caught definetely (e.g. the case for GRB
021004 and GRB 021211). However, even in this case, 
one can always define a
``pseudo reverse shock peak'' by recording the very first data
point in the observed $\propto t^{-2}$ lightcurve. Denoting this
point as $(\bar t_{p,r}, \bar F_{\nu,p,r})$, one can similarly
define \be \bar {\cal R}_t \equiv \frac{t_{p,f}}{\bar t_{p,r}}
\leq {{\cal R}_t}, ~~~ \bar {\cal R}_F \equiv \frac{\bar
F_{\nu,p,r}}{F_{\nu,p,f}} \leq {{\cal R}_F}. \label{bar} \ee
Repeating the above procedure, one can derive $\bar{\hat\gamma}$ 
and $\bar{\cal R}_\nu$ using (\ref{zeta1}) and (\ref{zeta2}),
respectively, but with parameters ${{\cal R}_t}$ and ${{\cal
R}_F}$ substituted by their ``bar'' counterparts. The
$\bar{\hat\gamma}$ and $\bar{\cal R}_\nu$ values, however, are
only upper (or lower) limits for $\hat\gamma$ and ${\cal R}_\nu$
for ${\cal R}_\nu>1$ (or ${\cal R}_\nu<1$). Usually we may be able
to estimate $t_{p,r}$, and hence ${{\cal R}_t}$, from other
constraints (e.g. $t_{p,r}$ has to be larger than $T$, etc.). In
such cases, we can derive $\hat\gamma$ and ${\cal R}_\nu$ from
$\bar{\hat\gamma}$ and $\bar {\cal R}_\nu$ with some correction
factors involving $(\bar {\cal R}_t/{{\cal R}_t})\leq 1$.  Using
eqs. (\ref{alpha}), (\ref{ft}), (\ref{fF}), (\ref{bar}), and
noticing $\bar {\cal R}_F/{{\cal R}_F}=(\bar {\cal R}_t/{{\cal
R}_t})^\alpha$ (as derived from $F_{\nu,r}\propto t^{-\alpha}$),
we have, for ${\cal R}_\nu > 1$,
\ba
\hat\gamma & = & \bar{\hat\gamma}
\left(\frac{\bar {\cal R}_t}{{{\cal
R}_t}}\right)^{\frac{3}{(4\alpha-7)}} \leq \bar{\hat\gamma}, \\
{\cal R}_\nu & = & \bar{\cal R}_\nu \left(\frac{\bar {\cal R}_t}{{{\cal
R}_t}}\right)^{\frac{3(4\alpha-3)}{2(4\alpha-7)}} \leq  \bar{\cal
R}_\nu ,
\ea
and for ${\cal R}_\nu < 1$,
\ba
\hat\gamma &
= & \bar{\hat\gamma} \left(\frac{\bar{\cal R}_t} {{\cal
R}_t}\right)^{-\frac{3}{4}} \geq \bar{\hat\gamma}, \\
{\cal R}_\nu & = & \bar{\cal R}_\nu \left(\frac{\bar{\cal R}_t}{ {\cal
R}_t}\right)^{-\frac{12\alpha-9}{8}} \geq \bar{\cal R}_\nu.
\ea
With an estimated $\hat\gamma$, one can again estimate $\gamma_0$
using (\ref{eta}). Notice that, however, unless one catches $t_{p,r}$,
both solutions for ${\cal R}_\nu >1$ and ${\cal R}_\nu<1$ are possible 
and one needs additional information to break the degeneracy.
To do this, some knowledge (but not the precise values) of the
microphysics parameters is usually needed. With eqs.(\ref{tcross}),
(\ref{fnum}), (\ref{zeta}), and eq.(1) of Kobayashi \& Zhang (2003a),
one can derive
\be
{\cal R}_\nu \sim 500 {{\cal R}_B}^{-1}
\gamma_{0,2}^{-2}n^{-1/2} \epsilon_{B,-2}^{-1/2} \epsilon_{e,-1}^{-2}
\left(\frac{g}{1/3}\right)^{-2} \frac{(1+z)}{2},
\ee
where $g=(p-2)/(p-1)$. We can see that generally ${\cal R}_\nu>1$, but
the ${\cal R}_\nu \siml 1$ case is also allowed for some extreme
parameters.

There are two caveats about our method. First, it involves a value of
${\cal R}_B$ (the reverse to forward comoving magnetic field ratio),
so one cannot determine $\gamma_0$ without specifying ${\cal
R}_B$. The usual standard assumption is that ${\cal R}_B=1$; however, 
it may be $>1$, e.g. if the central engine is strongly magnetized.
We note that there is another independent way to constrain
$\gamma_0$. Generally if $t_\times$ can be measured
one can directly derive (Sari \& Piran 1999b)
\be
\gamma_0 \geq \gamma_\times =\gamma_c \left(\frac{T}{t_\times}\right)^{3/8},
\label{gamcross}
\ee
which gives the value (or a lower limit) of $\gamma_0$ for the thin or
thick shell case, respectively. This is the case for GRB 990123. When
such independent information about $\gamma_0$ is available,  constraints
on ${\cal R}_B$ may be obtained (see \S5 for discussion of GRB 990123).

Second, in the above treatment we have adopted the strict adiabatic
assumption so that $F_{\nu,m,f}$ stays constant from $t_\times$ all
the way to $t_{p,f}$. In principle, radiation loss during the early
forward shock evolution may be important and ought to be taken into
account. This introduces a correction factor $f_{rad}\sim
(t_{p,f}/t_\times)^{(17/16)\epsilon_e}$ (Sari 1997), which depends on
$\epsilon_e$. Suppose the measured forward shock peak flux is
$F_{\nu,p,f}$(obs), then the real value of $F_{\nu,p,f}$ to be
used in the above method (e.g. to derive ${\cal R}_F$, eq.[\ref{fF}])
should be $F_{\nu,p,f}=f_{rad}F_{\nu,p,f}$(obs). For typical values of
$\epsilon_e$, e.g., $\sim 0.1$, typically we have $f_{rad}<2$.
Nonetheless, in detailed case studies, this correction factor
should be taken into account.

\section{Classification of early optical afterglows}

Regardless of the variety of early lightcurves, the reverse shock
$F_{\nu,r}\propto t^{-2}$ emission component is expected to eventually
join with the forward shock emission lightcurve. We can identify two
cases (Fig.1). {\em Type I (Rebrightening):} The reverse shock component
meets the forward shock component before the forward shock peak time, as
might have been observed in GRB 021004 (Kobayashi \& Zhang 2003a).
{\em Type II (Flattening):} The reverse shock component meets the forward
shock component after the forward shock peak time. GRB 021211 may be a
marginal such case (Fox et al. 2003; Li et al. 2003; Wei 2003).

The condition for a flattening type is $F_{\nu,r}(t_{p,f}) > F_{\nu,p,f}$.
Using eqs.(\ref{ft}) and (\ref{fF}), noticing $F_{\nu,r} \propto
t^{-\alpha}$, the flattening or type-II condition turns out to be
\be
\hat\gamma < {{\cal R}_B}^{\frac{2\alpha+3}{4\alpha-3}} {{\cal
R}_\nu}^{\frac{2}{4\alpha-3}} \sim {{\cal R}_B}^{7/5}
{{\cal R}_\nu}^{2/5}
\label{II}
\ee
for both ${\cal R}_\nu>1$ and ${\cal R}_\nu<1$, where the final
expression is for the typical value $\alpha=2$. We can see that the
type II condition is very stringent, especially when ${\cal R}_B$ is
not much above unity. We expect that rebrightening lightcurves should
be the common situation, unless the GRB central engines are typically
strongly magnetized. When a flattening lightcurve is observed, it is
likely that ${\cal R}_\nu \gg 1$, i.e, the peak frequency for the
reverse shock emission is well below the optical band. A very large
${\cal R}_\nu$ should usually involve very low luminosities in both
the reverse and forward shock emission. This might be the case of GRB
021211, which could have been categorized as an ``optically dark''
burst if the early reverse shock emission had not been
caught. Alternatively, a flattening case may be associated with a
strongly magnetized central engine, since a higher ${\cal R}_B$ can
significantly ease the type II condition. Finally, a large radiative
loss (with $\epsilon_e$ close to unity) may also at least contribute
to a flattening lightcurve.

\section{Case studies}

Early optical afterglows have been detected from GRB 990123, GRB
021004 and GRB 021211. Unfortunately, none of these observations give 
us an ``idealized'' data set, i.e., none of these cases showed
two distinctly identifiable peaks. Our method therefore cannot be 
straightforwardly applied to them. Nonetheless, we can use the
correction factors mentioned in \S3 to estimate $\gamma_0$ in these
cases. We expect that {\em Swift} will provide ideal data sets for
more bursts on which our method could be utilized directly.

1. GRB 990123: The basic parameters of this burst
include\footnote{Hereafter we assume that the kinetic energy of the
fireball in the deceleration phase is comparable to the energy
released in gamma-rays in the prompt phase.}  (e.g. Kobayashi \& Sari
2000 and references therein) $E_{52} \sim 140$, $z=1.6$, $T_2 \sim
0.6$, $\alpha \sim 2$.
This gives $\gamma_c \sim 305 n^{-1/8}$ (notice weak dependence on
$n$). The reverse shock peak was well determined: $(t_{p,r},
F_{\nu,p,r})\sim (50{\rm s}, 1{\rm Jy})$. The lightcurve shows ${\cal
R}_\nu > 1$. Since the early rising ligtcurve is caught, $t_\times$ is
directly measured, i.e., $t_\times \sim t_{p,r}
\sim T$. This is a marginal case, and $\gamma_0 \sim \gamma_c \sim
300 n^{-1/8}$ (eq.[\ref{gamcross}]). Unfortunately, the forward shock
peak was not caught. Assuming $t_{p,f}\sim 0.1$ d, one has ${{\cal
R}_t} \sim 170$, and ${{\cal R}_F} \sim 5000$. The radiative
correction factor is $f_{rad}\sim 2$ by adopting $\epsilon_e \sim 0.13$
(Panaitescu \& Kumar 2001). With this correction, one has $\hat\gamma
\sim (0.07 {{\cal R}_B}^{5/3})^{3}$ (eq.[\ref{zeta1}]). By requiring
$\hat\gamma \sim 300 n^{-1/8}$, ${\cal R}_B \sim 15n^{-1/40}$ is
required. We conclude that GRB 990123 is giving us the first evidence
for a strongly magnetized central engine.

2. GRB 021004: The parameters of this burst are (e.g. Kobayashi \&
Zhang 2003a and
references therein) $E_{52} \sim 5.6$, $z=2.3$, $T_2 \sim 1$. We have
$\gamma_c \sim 190 n^{-1/8}$. The forward shock peak is reasonably
well measured, but the reverse shock peak is not caught. Using the
first data point as modeled in Kobayashi \& Zhang (2003a), we have
$\bar {\cal R}_t \sim
12$, $\bar {\cal R}_F \sim 2$. Solving for $\hat\gamma$ and ${\cal
R}_\nu$, we get $\hat \gamma_0 \leq 220 {{\cal R}_B}^5$ for ${\cal
R}_\nu > 1$ and $\alpha \sim 2$ (in the
asympototic phase), and $\hat\gamma \geq 6.5 {{\cal R}_B}^{1/2}$ for
${\cal R}_\nu<1$. The detailed modeling of the lightcurve suggests
${\cal R}_\nu > 1$ and {\bf a} thin shell, so that $\gamma_0 \sim 120$ for
${\cal R}_B \sim 1$ (Kobayashi \& Zhang 2003a).

3. GRB 021211: The basic parameters of this burst include $E_{52} \sim
0.6$, $z=1.0$, $T_2\sim 0.15$, $\alpha \sim 1.8$ (Li et al. 2003; Fox
et al. 2003; and references therein), so that $\gamma_c \sim 240
n^{-1/8}$. Neither the forward nor the reverse shock peaks are
identified. Assuming that the forward shock peak occurs close to the
flattening break (marginal type-II flattening case), and taking the
first data point of the early lightcurve, we get $\bar {\cal
R}_t\sim 20$, $\bar {\cal R}_F\sim 30$. Since ${\cal R}_\nu<1$ is
unlikely for a flattening (type-II) lightcurve (eq.[\ref{II}]), we get
$\hat\gamma \leq (0.4{{\cal R}_B}^{1.5})^{15}$ with ${\cal R}_\nu>1$.
For any reasonable $\hat\gamma$, ${\cal R}_B$ has to be somewhat
larger than unity. More information would be needed in order
to estimate $\hat\gamma$ and $\gamma_0$.

\section{Conclusions and implications}

We have discussed a straightforward recipe for constraining the 
initial Lorentz factor $\gamma_0$ of GRB fireballs by making use of
the early optical afterglow data alone. Data in other bands
(e.g. X-ray or radio) are generally not required. The input parameters
are ratios of observed emission quantities, so that poorly known model
parameters related to the shock microphysics (e.g. $\epsilon_e$,
$p$, etc.) largely cancel out if they are the same in both
shocks. Otherwise, our method only invokes ratios of these parameters
(e.g. ${\cal R}_B$, eq.[\ref{b}]) and the absolute values of the
microphysics parameters do not enter the problem. This approach is
readily applicable in the {\em Swift} era when many early optical
afterglows are expected to be regularly caught.

Regular determinations of the initial Lorentz factors of a large
sample of GRBs would have profound implications for our understanding
of the nature of bursts. Currently only lower limits on $\gamma_0$ are
available for some bursts (e.g. Baring \& Harding 1997; Lithwick \&
Sari 2001). Since in various models
some crucial GRB parameters (such as the energy peak in the spectrum)
depend on the fireball Lorentz factor in different ways (see Zhang \&
\Mesz~2002b for a synthesized study), a large sample of $\gamma_0$ data
combined with other information (which is usually easier to be
attained in the {\em Swift} era) may allow us to identify the location
and mechanism of the GRB prompt emission through statistical analyses. 
Searching for the possible correlation between the GRB prompt emission 
luminosities and Lorentz factors can allow tests for the possible GRB 
jet structures proposed in some models (e.g. Zhang \& \Mesz~2002a).

We have also classified the early optical afterglow lightcurves into
two types. The rebrightening case (type-I) is expected to be
common. The flattening case (type-II) may be rare, and when detected,
is likely to involve a low luminosity or a strongly magnetized central
engine. 

It is worth emphasizing that there is evidence that the
central engine of GRB 990123 is strongly magnetized. A similar
conclusion may also apply to GRB 021211. If from future data a
strongly magnetized ejecta is commonly inferred, this would stimulate
the GRB community to seriously consider the role of strong magnetic
fields on triggering the GRB prompt emission as well as collimating
the GRB jets. Recently, the gamma-ray prompt emission of a bright
burst, GRB 021206, was detected by the {\em RHESSI} mission and
reported to be strongly polarized (Coburn \& Boggs 2003). This can be 
interpreted as evidence for a strongly magnetized ejecta. Our method
leads to an independent and similar conclusion for another bright
burst, GRB 990123. We also note that, since a strong reverse shock
component is detected, the fireball could not be completely
Poynting flux dominated (which should have greatly suppressed the
reverse shock emission, Kennel \& Coroniti 1984). In other words,
the fireball does not have to be in the high-$\sigma$ regime (see
Zhang \& \Mesz~ 2002b for a discussion of various fireball regimes). 
A likely picture is a kinematic-energy-dominated fireball entrained
with a strong (maybe globally organized) magnetic component.

\acknowledgements This work is supported by NASA NAG5-9192, NAG5-9153
and the Pennsylvania State University Center for Gravitational Wave
Physics, which is funded by NSF under cooperative agreement PHY
01-14375.

\begin{figure}
\plotone{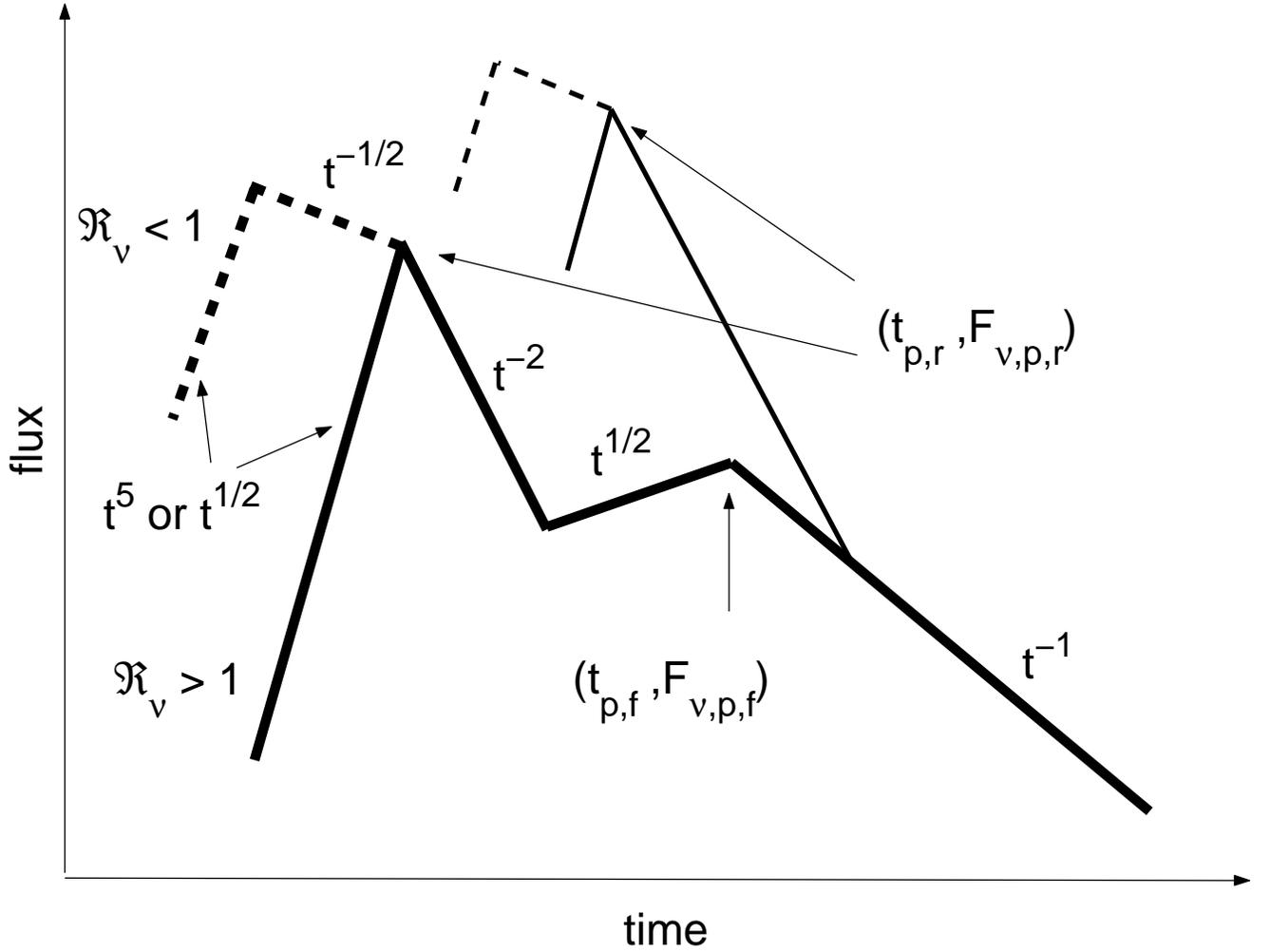} \caption{ Typical lightcurves of the
reverse-forward shock emission combinations for the homogeneous ISM
case. The thick lines
depict a typical ``rebrightening'' (type I) lightcurve, while thin
lines indicate a typical ``flattening'' (type II) lightcurve. The
forward shock peak $(t_{p,f}, F_{\nu,m,p})$ is defined at the
transition point of the $\propto t^{1/2}$ to $\propto t^{-1}$
lightcurves. The reverse shock peak $(t_{p,r}, F_{\nu,m,r})$ is
defined at the beginning of the $\propto t^{-2}$ segment for the
reverse shock emission. Before this point, the lightcurve is
$\propto t^5$ (thin shell) or $\propto t^{1/2}$ (thick shell) for
${\cal R}_\nu\equiv \nu_R/\nu_{m,r}(t_\times) >1$ (usually the
case), and $\propto t^{-1/2}$ for ${\cal R}_\nu <1$.
 \label{fig1}}
 \end{figure}


\begin{references}
\reference{} Akerlof, C. W. et al. 1999, Nature, 398, 400
\reference{} Baring, M. G. \& Harding, A. K. 1997, ApJ, 491, 663
\reference{} Chevalier, R. A. \& Li, Z.-Y. 1999, ApJ, 520, L29
\reference{} Coburn, W. \& Boggs, S. E. 2003, Nature, 423, 415
\reference{} Crew, G. et al. 2002, GCN 1734,
(http://gcn.gsfc.nasa. gov/gcn/gcn3/1734.gcn3)
\reference{} Fan, Y. Z., Dai, Z. G., Huang, Y. F. \& Lu, T. 2002,
ChJAA, 2, 449
\reference{} Fox, D. W. 2002, GCN 1564
(http://gcn.gsfc.nasa.gov/gcn/gcn3/ 1564.gcn3)
\reference{} Fox, D. W. \& Price, P. A. 2002, GCN 1731
(http://gcn.gsfc.nasa. gov/gcn/gcn3/1731.gcn3)
\reference{} Fox, D. W. et al. 2003, ApJ, 586, L5
\reference{} Kennel, C. F., \& Coroniti, F. V. 1984, ApJ, 283, 694
\reference{} Kobayashi, S 2000, ApJ, 545, 807
\reference{} Kobayashi, S, Piran, T \& Sari, R. 1999, ApJ, 513, 669
\reference{} Kobayashi, S, \& Sari, R. 2000, ApJ, 542, 819
\reference{} Kobayashi, S, \& Zhang, B. 2003a, ApJ, 582, L75
\reference{} -----. 2003b, ApJ, submitted
(astro-ph/0304086)
\reference{} Li, W., et al., 2002,
GCN 1737 (http://gcn.gsfc.nasa.gov/gcn/gcn3/ 1737.gcn3)
\reference{} Li, W., et al. 2003, ApJ, 586, L9
\reference{} Lithwick, Y. \& Sari, R. 2001, ApJ, 555, 540
\reference{} \Mesz, P. \& Rees,M.J. 1997a, ApJ, 476, 231
\reference{} -----. 1997b, ApJ, 482, L29
\reference{} -----. 1999, MNRAS, 306, L39
\reference{} Panaitescu, A., \& Kumar, P. 2001, ApJ, 560, L49
\reference{} Park, H. S., Williams, G. \& Barthelmy, S. 2002, GCN 1736\\
(http://gcn.gsfc.nasa. gov/gcn/gcn3/1736.gcn3)
\reference{} Sari, R. 1997, ApJ, 489, L37
\reference{} Sari, R. \& Piran, T. 1995, ApJ, 455, L143
\reference{} -----. 1999a, ApJ, 517, L109
\reference{} -----. 1999b, ApJ, 520, 641
\reference{} Sari, R., Piran, T. \& Narayan, R. 1998, ApJ, 497, L17
\reference{} Shirasaki, Y. et al. 2002, GCN 1565
(http://gcn.gsfc.nasa. gov/gcn/gcn3/1565.gcn3)
\reference{} Soderberg, A. M., \& Ramirez-Ruiz, E. 2002, MNRAS, 330, L24
\reference{} Usov, V. V. 1992, Nature, 357, 472
\reference{} Wang, X. Y., Dai, Z. G., \& Lu, T. 2000, MNRAS, 319, 1159
\reference{} Wei, D. M. 2003, A\&A, 402, L9
\reference{} Wheeler, J. C., Yi, I., H\"oflich, P. \& Wang, L. 2000,
ApJ, 537, 810
\reference{} Wolzniak, P., et al. 2002, GCN 1757
(http://gcn.gsfc.nasa. gov/gcn/gcn3/1757.gcn3)
\reference{} Zhang, B. \& \Mesz, P. 2002a, ApJ, 571, 876
\reference{} -----. 2002b, ApJ, 581, 1236

\end{references}
\end{document}